\documentclass[aps,prl,amsmath,amssymb,reprint,longbibliography,superscriptaddress,
preprintnumbers]{revtex4-1}
\usepackage{graphicx}  
\usepackage{dcolumn}   
\usepackage[]{natbib}
\usepackage{color}
\usepackage{times}
\usepackage{hyperref}

\newcommand{\be}{\begin{equation}}
\newcommand{\ee}{\end{equation}}
\newcommand{\ba}{\begin{eqnarray}}
\newcommand{\ea}{\end{eqnarray}}

\newcommand{\nn}{\nonumber\\}

\def\pa{\partial}

\def\a{\alpha}
\def\b{\beta}

\def\G{\Gamma}

\def\D{\Delta}
\def\e{\epsilon}

\def\l{\lambda}

\def\m{\mu}

\def\s{\sigma}

\def\t{\tau}

\def\be{\begin{eqnarray}}
\def\ee{\end{eqnarray}}
\def\D{\Delta}
\def\t{\tau}
\def\a{\alpha}
\def\b{\beta}
\def\m{\mu}

\def\D{\Delta}
\def\l{\lambda}

\def\G{\Gamma}

\def\nn{\nonumber\\}
\def\pa{\partial}
\def\s{\sigma}
\def\e{\epsilon}

\newcommand\<\langle
\renewcommand\>\rangle

\begin{document}

\title{Ising Model Close to $d=2$}

\author{Wenliang Li}
\email{liwliang3@mail.sysu.edu.cn}
\affiliation{School of Physics, Sun Yat-Sen University, Guangzhou 510275, China}
\affiliation{Okinawa Institute of Science and Technology Graduate University, 1919-1 Tancha, Onna-son, Okinawa 904-0495, Japan}

\begin{abstract}
The $d=2$ critical Ising model is described by an exactly solvable Conformal Field Theory (CFT). 
The deformation to $d=2+\epsilon$ is a relatively simple system at strong coupling outside of even dimensions.  
Using novel numerical and analytical conformal bootstrap methods in Lorentzian signature, 
we show that the leading corrections to the Ising data are more singular than $\e$. 
There must be infinitely many new states due to the $d$-dependence of conformal symmetry. 
The linear independence of conformal blocks is central to this bootstrap approach, 
which can be extended to more rigorous studies of non-positive systems, 
such as non-unitary, defect/boundary and thermal CFTs. 
\end{abstract}

\maketitle 
\section{Introduction}
The $d$-dimensional Ising model is a fundamental model in statistical physics and condensed matter physics. 
Historically, it was proposed by Lenz to describe ferromagnetism and the case of $d=1$ was solved by Ising. 
This simple model displays rich physics and captures some main traits of phase transitions and many-body problems. 
At criticality, it belongs to one of the simplest universality classes, characterized by the global $\mathbb Z_2$ symmetry. 
For $d>4$, the critical behaviour of the Ising model is described by Landau's mean-field theory \cite{Landau:1937obd}, 
in which fluctuations are neglected due to the averaging effects of many adjacent spins.  
At lower $d$, fluctuations play a more significant role.  
The mean-field description is not sufficient for $d\leq 4$  
and the Ising critical exponents have non-trivial $d$-dependence \cite{fn1}. 
As a natural continuation of Landau's theory, 
Wilson and Fisher calculated the critical exponents in $d=4-\epsilon$ dimensions 
using the perturbative $\epsilon$ expansion \cite{Wilson:1971dc}. 
The $\epsilon$ expansion has proved to be a valuable tool in the studies of critical phenomena \cite{Wilson:1973jj,Pelissetto:2000ek}.

At $d=2$, it is well-known that the Ising model is solvable 
since Onsager's groundbreaking results \cite{Onsager:1943jn}. 
The critical behaviour is described by the fixed point of renormalization group flows. 
In particular, scale invariance of the fixed point is promoted to conformal invariance. 
As another natural continuation, it would be interesting to deform the $2d$ exact solution to $d=2+\epsilon$ dimensions. 
The $\epsilon$ expansion usually concerns weakly coupled systems 
\cite{Migdal:1975zf,Polyakov:1975rr,Brezin:1975sq,Brezin:1976qa}, 
but the case here remains strongly coupled \cite{fn4}, 
so the intriguing strong coupling physics becomes more manifest. 
More recently, the $\e=d-2$ expansion has also been used to study deconfined quantum criticality  \cite{Ma:2019ysf,Nahum:2019fjw}. 
(See \cite{He:2020azu} for a numerical conformal bootstrap study.) 
We notice a deceptively simple question:

{\it Is the $\e$ expansion of a strongly coupled system given by integer power series?} 

In the standard $\e$ expansion, 
the corrections to the $d=4$ data can be computed order by order in $\epsilon$, 
given by asymptotic series \cite{Brezin:1976vw}. 
It has been argued that they are integer power series     
based on the Renormalization Group (RG) analysis in the minimal subtraction scheme \cite{Bagnuls-Bervillier, Schafer}. 
Naively, one might think that the $\e=d-2$ expansion should also be the case. 
For instance, 
the scaling dimension of the lowest $\mathbb Z_2$-even operator was assumed to 
receive integer power corrections in the study of disorder effects in $2+\e$ dimensions \cite{Komargodski:2016auf}. 
However, the standard $\e$ expansion is around a Gaussian theory. 
The weak coupling techniques and arguments do not easily extend to the strong coupling situation. 
For the $O(n)$ model, Cardy and Hamber performed an elegant analysis around $n=d=2$ 
based on some analyticity assumptions on the RG equations \cite{Cardy:1980at}, 
but these results do not apply for $d=2+\e$ with $n<2$. 

In this letter, we will study the $d=2+\e$ Ising model using the conformal bootstrap. 
The conformal bootstrap program aims to classify and solve CFTs by general principles and consistency conditions 
\cite{Ferrara:1973yt,Polyakov:1974gs}, without resorting to the weak coupling expansion.  
For $d=2$, conformal symmetry becomes infinite-dimensional  
and this program can be carried out rather successfully \cite{Belavin:1984vu,DiFrancesco:1997nk}. 
The studies in $d>2$ dimensions are more challenging as conformal symmetry is less constraining. 
Nevertheless, considerable progress has been achieved due to the seminal work \cite{Rattazzi:2008pe}, 
in which the unitarity assumption and the crossing equations are formulated as inequalities. 
This modern bootstrap approach has led to rigorous bounds on the space of unitary CFTs, 
such as the most precise determinations of the $d=3$ Ising critical exponents 
\cite{ElShowk:2012ht, El-Showk:2014dwa, Kos:2014bka, Kos:2016ysd}.    
We refer to \cite{Qualls:2015qjb, Rychkov:2016iqz, Simmons-Duffin:2016gjk, Poland:2018epd, Chester:2019wfx} for useful reviews and lecture notes. 

The critical Ising model can be viewed as a continuous family of $\mathbb Z_2$-covariant CFTs parametrized by $d$. 
The case of non-integer $d$ has also been studied by the unitary bootstrap methods in 
\cite{El-Showk:2013nia,Behan:2016dtz,Cappelli:2018vir}. 
The bounds exhibit similar features as those at $d=2,3$ and the results are consistent with the $(4-d)$ expansion. 
However, a subtlety is that the Wilson-Fisher fixed point is non-unitary in non-integer dimensions, 
because the spectrum contains descendant states of complex scaling dimensions \cite{Hogervorst:2015akt}. 
It would be helpful to consider complementary approaches that are not based on unitarity, 
such as the flow method \cite{El-Showk:2016mxr} and the truncation method \cite{Gliozzi:2013ysa}. 
The truncated bootstrap approach has been applied to the study of non-positive problems 
\cite{Gliozzi:2014jsa,Gliozzi:2015qsa,Nakayama:2016cim,Esterlis:2016psv, Gliozzi:2016cmg,Hikami:2017hwv,Hikami:2017sbg,Hikami:2018mrf,LeClair:2018edq,
Hikami:2018qpz, Rong:2020gbi, Nakayama:2021zcr}. 
In the original formulation \cite{Gliozzi:2013ysa}, the truncated problem is encoded in determinants. 
In \cite{Li:2017ukc}, we proposed some new ingredients, 
which we believe are important to a more systematic formulation.   
We emphasized the essential role of linear independence and 
introduced the concept of norm to the truncation approach. 
(See \cite{Kantor:2021kbx,Kantor:2021jpz} for the recent implementation using reinforcement-learning algorithms.)
We will apply these notions to the numerical bootstrap study of the $d=2+\e$ Ising CFT. 

On the other hand, it was noticed in \cite{Cappelli:2018vir} that the tentative spectrum from the unitary numerical approach   
exhibits a transition at $d=2+\e$ with $\e\sim 0.2$ small but finite. 
Such a transition is expected since $d=2$ is special.   
At $d=2$, the spectrum is organized into Virasoro multiplets and 
the corresponding Regge trajectories have constant twists $\t=\D-\ell$ with integer spacing. 
At $d=3$, the twist spectrum of the Ising CFT is additive and 
the Regge trajectories have more interesting dependence on spin. 
Infinitely many high spin operators have twists asymptotic to the sum of two lower twists 
\cite{Fitzpatrick:2012yx, Komargodski:2012ek}.  
For example, the Regge trajectories $[\s\s]_n$ are associated with the lowest $\mathbb Z_2$-odd scalar $\s$ 
and they have twist accumulation points at $2\D_\s+2n$. 
We will discuss the location of the transition to the double-twist spectrum using analytic bootstrap techniques.  

To address our question, we will study the 4-point function of  
the lowest $\mathbb Z_2$-odd scalar operator $\s$. 
We focus on the leading corrections and assume: 
\begin{enumerate}
\item
The critical Ising model is conformally invariant in $d=2+\e$ dimensions with $|\e|\ll 1$.  
\item
The leading corrections to the $2d$ data are linear in $\epsilon$:
\be
\D_\s&=&\D_\s^{(0)}+\e\,\D_\s^{(1)}+\dots\,,
\\
\D_i&=&\D_i^{(0)}+\e\,\D_i^{(1)}+\dots\,,
\\
\l_i&=&\l_i^{(0)}+\e\,\l_i^{(1)}+\dots\,,
\ee
where $\D_i=\D_{\mathcal O_i}$ and $\l_i=\l_{\s\s\mathcal O_i}$ are the scaling dimension and OPE coefficient of $\mathcal O_i$. 
The zeroth order values can be derived from the exact solution at $d=2$. 
We will further assume that $\D_i^{(1)},\,\l_i^{(1)}$ do not grow too rapidly with $\D_i^{(0)}$, 
so the conformal block summation is convergent \cite{fn5}. 
\end{enumerate}
In the first assumption, scale invariance of the Ising fixed point is enhanced to conformal invariance. 
There is ample evidence for conformal invariance in $d=2, 3, 4-\epsilon$ dimensions, 
so we expect that this property extends to $d=2+\e$ \cite{fn6}. 
In the second assumption, the leading corrections cannot be more singular 
since they have positive integer powers. 
They cannot start from second or higher orders in $\e$
because the $d$-dependence of conformal blocks leads to first-order terms in the crossing equation. 

Below we will examine if this is a consistent scenario. 
It turns out that the assumptions 1 and 2 are not consistent, 
so the corrections are expected to be more singular than $\e^1$. 

\section{The crossing equation}
We consider the 4-point function of the lowest $\mathbb Z_2$-odd operator $\s$:
\be
\langle \s(x_1)\,\s(x_2)\,\s(x_3)\,\s(x_4)\rangle
=
\frac {\mathcal G(z,\bar z)} {x_{12}^{2\D_\s} x_{34}^{2\D_\s}}\,.
\ee
The conformally invariant cross-ratios are
\be
u=z\bar z=\frac {x_{12}^2\, x_{34}^2}  {x_{13}^2\, x_{24}^2},\quad
v=(1-z)(1-\bar z)=\frac {x_{14}^2\, x_{23}^2}  {x_{13}^2\, x_{24}^2}\,.
\ee
The crossing equation for $\mathcal G(z,\bar z)$ reads:
\be
v^{\D_\s} \mathcal G(z,\bar z)=u^{\D_\s} \mathcal G(1-\bar z,1-z)\,.
\ee
In the $\epsilon=d-2$ expansion, we have
\be
\mathcal G(z,\bar z)=\mathcal G^{(0)}(z,\bar z)
+\e\,\mathcal G^{(1)}(z,\bar z)+\dots\,,
\label{BigG-esp}
\ee
where the 2d solution reads
\be
\mathcal G^{(0)}(z,\bar z)=
\frac{\sqrt{1+\sqrt{u}+\sqrt{v}}}{\sqrt 2\,v^{1/8}}\,,
\label{G0}
\ee
and $\mathcal G^{(1)}(z,\bar z)$ can be written as convergent power series in $z,\bar z$ 
in the regime $0\le z,\bar z<1$. 
After the conformal block decomposition, the crossing equation becomes
\be
\sum_i \l_i^2\, F_i(z,\bar z)=0\,,
\ee
where
$
F_i(z,\bar z)=v^{\D_\s} G_{\D_i,\ell_i}(z,\bar z)-(z\leftrightarrow 1-\bar z)$  
and $G_{\D_i,\ell_i}$ is the global conformal block for the conformal multiplet 
labelled by the primary operator $\mathcal O_i$.  
To first order in $\e$, the crossing equation reads
\be
&&\sum_i\l_i^{(0)}\big(\l_i^{(0)}\D_\s^{(1)}\pa_{\D_\s}+\l_i^{(0)}\D_i^{(1)}\pa_{\D_i}
+2\, \l_i^{(1)}\big) F_i(z,\bar z)
\nn&=&
(-1)\sum_i\,\l_i^2\, \pa_d F_i(z,\bar z)\,,\quad
\label{crossing}
\ee
which will be written more compactly in \eqref{crossing-compact}. 
Note that the derivative $\pa_d$ extracts the $d$-dependence of $G_{\D,\ell}$. 
After taking the derivatives, we set $\{d,\D_i,\l_i\}\rightarrow \{2, \D_i^{(0)},\l_i^{(0)}\}$.
We do not make any assumptions about the signs of $\{\D_\s^{(1)},\,\D_i^{(1)},\,\l_i^{(1)}\}$.

Although the left hand side of \eqref{crossing} involves an infinite number of free parameters, 
the building blocks are the simple $2d$ conformal blocks 
\cite{fn7}
\be
G^{d=2}_{\D,\ell}(z,\bar z)=\frac{1}{1+\delta_{\ell,0}}\Big(k_{\D+\ell}(z)k_{\D-\ell}(\bar z)
+(z\leftrightarrow \bar z)\Big)\,,\qquad
\label{2d-block}
\ee
where $k_{\b}(x)=x^{\b/2}{}_2F_1(\b/2,\b/2,\b,x)$ is the $SL(2,\mathbb R)$ block with identical external scaling dimensions. 
Since each term is multiplied by $\l_i^{(0)}$, the intermediate states are the same as those in 2d and their twists are given by
\be
\{\t_i^{(0)}\}=\{4n, 4n+1\}\,,
\label{tau0}
\ee
where $n=0,1,2,\dotsb$ but $\t^{(0)}\neq 5$. 
Note that the twist-5 trajectory and the twist-1 spin-2 state are absent in the 2d intermediate spectrum of the $\s\times\s$ OPE. 
New states cannot contribute to the OPE at order $\e^1$ because their squared OPE coefficients are at least of order $\e^2$. 
This applies to both primary and descendant states 
\cite{fn8}.  
Since the 2d Ising model is unitary, 
we do not need to worry about potential cancellation of finite mixed contributions. 
On the contrary, 
the right hand side of \eqref{crossing} has no free parameter. 
We can compute the sum based on the 2d data 
using the general $d$ formula of conformal blocks \cite{Li:2019cwm}. 
Then we take the $d$ derivative and set $d\rightarrow 2$. 

In the ``bra-ket" notation, the crossing equation $\eqref{crossing}$ reads
\be
\D_\s^{(1)}|\D_\s\rangle+\sum_i\Big(\D_i^{(1)}|\D_i\rangle+\l_i^{(1)}|\l_i\rangle\Big)=-|d\rangle\,,
\label{crossing-compact}
\ee
where $|a\rangle$ denotes the contribution generated by the change in $a$. 
Our question in the introduction becomes: 
{\it
Do $|\D_\s\rangle, |\D_i\rangle, |\l_i\rangle$ form a complete set of basis for $|d\rangle$?
}
It turns out that the answer is negative \cite{fn9}! 
The target $|d\rangle$ does not belong to the vector space spanned by 
$\{|\D_\s\rangle, |\D_i\rangle, |\l_i\rangle\}$.

Before analyzing the crossing equation \eqref{crossing-compact}, 
let us discuss the building blocks $|\D_\s\rangle\,,|\l_i\rangle,\,|\D_i\rangle, |d\rangle$. 
The first one can be easily derived from 
\eqref{G0}. 
Then a global conformal block takes a factorized form at $d=2$, given in \eqref{2d-block}. 
(See \cite{Li:2020ijq} for a general $d$ generalization. )
According to the $z$ dependence,
we have 
\be
\label{k-sum}
&&\sum_i\Big(\D_i^{(1)}|\D_i\rangle+\l_i^{(1)}|\l_i\rangle\Big)
\\&=&
\sum_{\b} v^{\frac 1 8}\big(A_\b(\bar z)k_{\b}(z) +B_\b(\bar z)\pa_\b k_{\b}(z) \big)
-(z \leftrightarrow 1-\bar z)\,,\nonumber
\ee
where $\b\in\{\tau^{(0)}\}$ is defined in \eqref{tau0} and
$A_\b(\bar z), B_\b(\bar z)$ encode the dependence on $\bar z$.  
We can use the general $d$ formula in \cite{Li:2019cwm} to compute numerically 
$|d\rangle$ order by order in $z$ at any $\bar z$ in $[0,1)$ \cite{fn10}. 
The analytic computation of $|d\rangle$ based on \cite{Simmons-Duffin:2016wlq} is described in Supplemental Material.  

\section{Numerical conformal bootstrap}
Let us perform a numerical study of the crossing equation \eqref{crossing-compact}, 
which has no solution if $\{|d\rangle, |\D_\s\rangle,|\D_i\rangle,|\l_i\rangle\}$ are linearly independent. 
We can detect the linear independence by a norm: 
\be
\eta=\Big\Vert 
|d\rangle+
\D_\s^{(1)}|\D_\s\rangle+\sum_{i}\D_i^{(1)}|\D_i\rangle+\sum_{i} \l_i^{(1)}|\l_i\rangle
\Big\Vert\,,\qquad
\ee
which is the distance between the target point determined by $|d\rangle$ and 
a point in the space spanned by $|\D_\s\rangle,|\l_i\rangle,|\D_i\rangle$. 
If there exists at least one crossing solution, 
then we should find $\eta_\text{min}=0$. 
We define the norm in terms of sampling points
\cite{Hogervorst:2013sma} 
\be
\Vert H\Vert
=\sqrt{\langle H|H\rangle }
=\bigg(\frac 1 N \sum_{i=1}^N\,\mu(z_i,\bar z_i)\,\big| H(z_i,\bar z_i)\big|^2\bigg)^{1/2}
\,,\qquad
\ee
where the measure $\mu(z,\bar z)$ will be specified later. 
The inner product $\langle H_1|H_2\rangle$ is defined as a weighted sum of the product $H_1^\ast H_2$. 
We consider the Lorentzian regime, so $z,\bar z$ are independent, real variables. 
We further concentrate on the region near the double-lightcone limit with $0<z\ll 1$ and $0\ll\bar z<1$, 
which will also be studied analytically. 
We use sampling rather than derivative equations
because it is easier to assign a proper measure $\mu(z,\bar z)$. 

In practice, we need to truncate the conformal block summation to a finite sum in the numerical studies. 
This is sometimes called OPE truncation \cite{fn13}. 
Then we need to know if a finite $\eta_\text{min}$ is 
due to the OPE truncation or absence of crossing solution. 
Since we are sampling in a subregion, the prefactor of $\eta_\text{min}$ is scheme-dependent and 
the finite $\eta_\text{min}$ becomes smaller as we increase the truncation cutoff. 
To distinguish between the two origins, 
we examine the dependence of $\eta_\text{min}$ on the local sampling regions labelled by $z_0$. 
If $\eta_\text{min}>0$ is mainly due to the OPE truncation, 
then the functional form of $\eta_\text{min}(z_0)$ will change dramatically with the cutoff. 
Otherwise, $\eta_\text{min}(z_0)$ will only get a smaller prefactor 
as more intermediate states are introduced. 
Near the lightcone, 
we can readily distinguish between them based on the scaling behaviour. 

\begin{figure}[h!]
\begin{center}
\includegraphics[width=8.6cm]{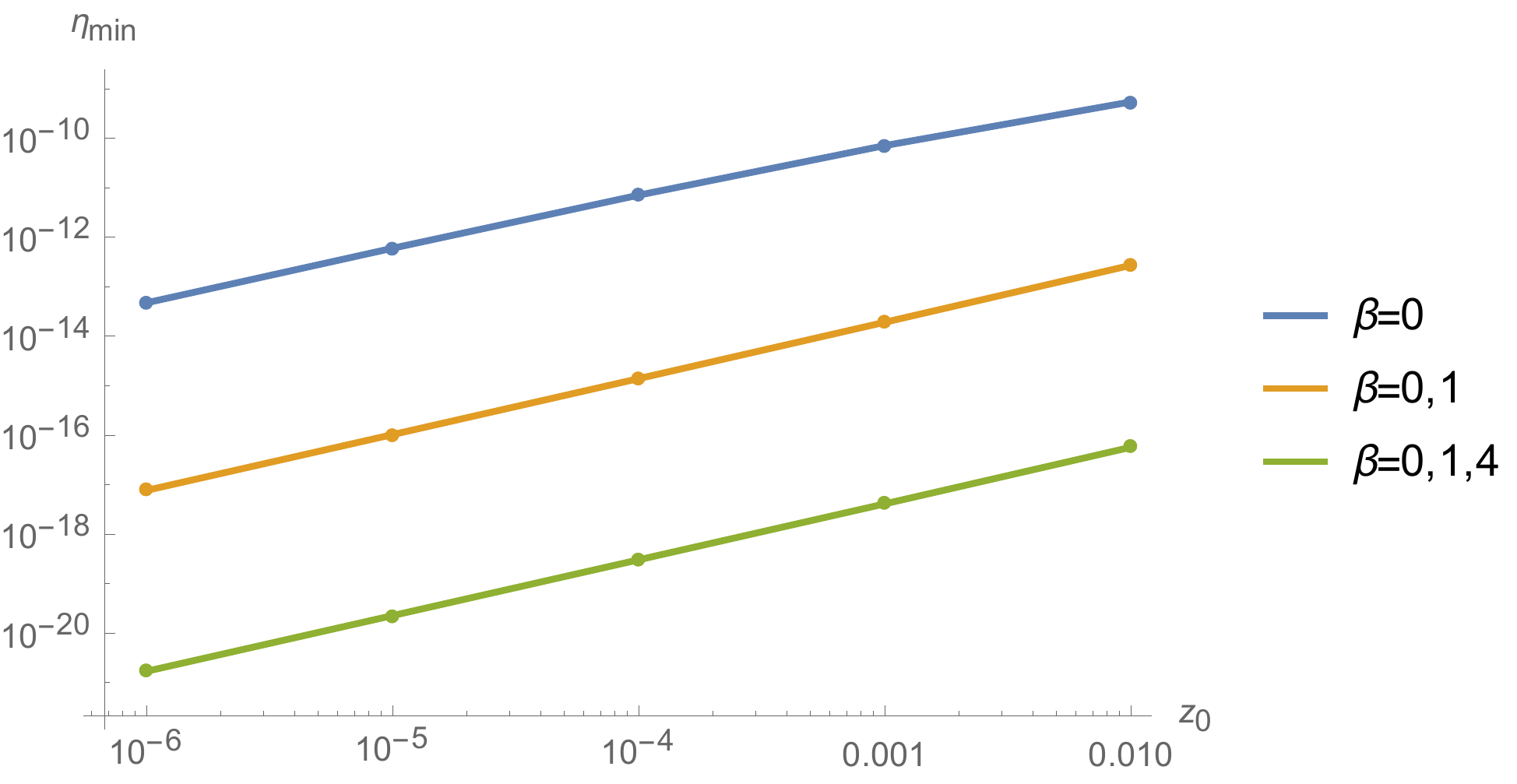}
\caption{Log-log plot of $\eta_\text{min}(z_0)$ with a simple measure \eqref{simple-mu},  
where $z_0$ labels the sample region. 
The sampling points are at  
 $z,1-\bar z=z_0\times 10^{-k/10}$ with $z\neq 1-\bar z$ and $k=0,1,2, \dots, 10$. 
The scaling behaviour is not sensitive to the $\b$ truncation. 
A larger $\b$ cutoff reduces the prefactor, but does not modify the leading scaling behaviour. 
Therefore, a finite $\eta_\text{min}$ is mainly due to the absence of crossing solution, not the $\b$ truncation. 
}
\label{figure:scaling}
\end{center}
\end{figure}

Near the lightcone $z=0$, we can truncate the sum \eqref{k-sum} to low $\b$. 
But we will not truncate the spin sum, 
so $A_\b(\bar z), B_\b(\bar z)$ in \eqref{k-sum} remain arbitrary \cite{fn14}.
They will be evaluated near the other lightcone $\bar z=1$. 
In the $\eta$ minimization, 
$A_\b(\bar z)$ and $B_\b(\bar z)$ are approximated by truncated Taylor series about 
the center of the sampling region. 
We use high order Taylor polynomials to make the associated errors negligible. 
We can also view $A_\b(\bar z), B_\b(\bar z)$ at different $\bar z$ as independent parameters,  
but the results remain the same in the cases examined. 

Now we discuss the choice of $\mu$. 
A simple measure is
\be
\mu_\text{simple}(z,\bar z)=1\,.
\label{simple-mu}
\ee
In Fig. \ref{figure:scaling}, we show the dependence of $\eta_\text{min}$ 
on the sampling region labelled by $z_0$. 
One can notice the scaling behaviour
\be
\eta_\text{min}(z_0)\propto z_0{}^{\a}\,,
\ee
which becomes more precise at small $z_0$. 
The exponent $\a$ is about $1.13(1)$ in the regime $10^{-6}<z_0<10^{-3}$, 
in which $|d\rangle$ can be computed by a direct summation over spin. 
For $z_0<10^{-6}$, we use the analytic expression of $|d\rangle$ in Supplemental Material 
to obtain a more precise value $\a\approx1.125$. 
These results imply that $|d\rangle$ contains a vector 
that scales as $\l^{1.125}$ under $\{z,1-\bar z\}\rightarrow \{ \l z, \l(1-\bar z)\}$.  
Furthermore, it does not belong to the space spanned by $\{|\D_\s\rangle\,,|\l_i\rangle,\,|\D_i\rangle\}$,  
so no crossing solution can be found. 
We will give an analytic understanding of the linear independence later. 

\begin{figure}[h!]
\begin{center}
\includegraphics[width=8.6cm]{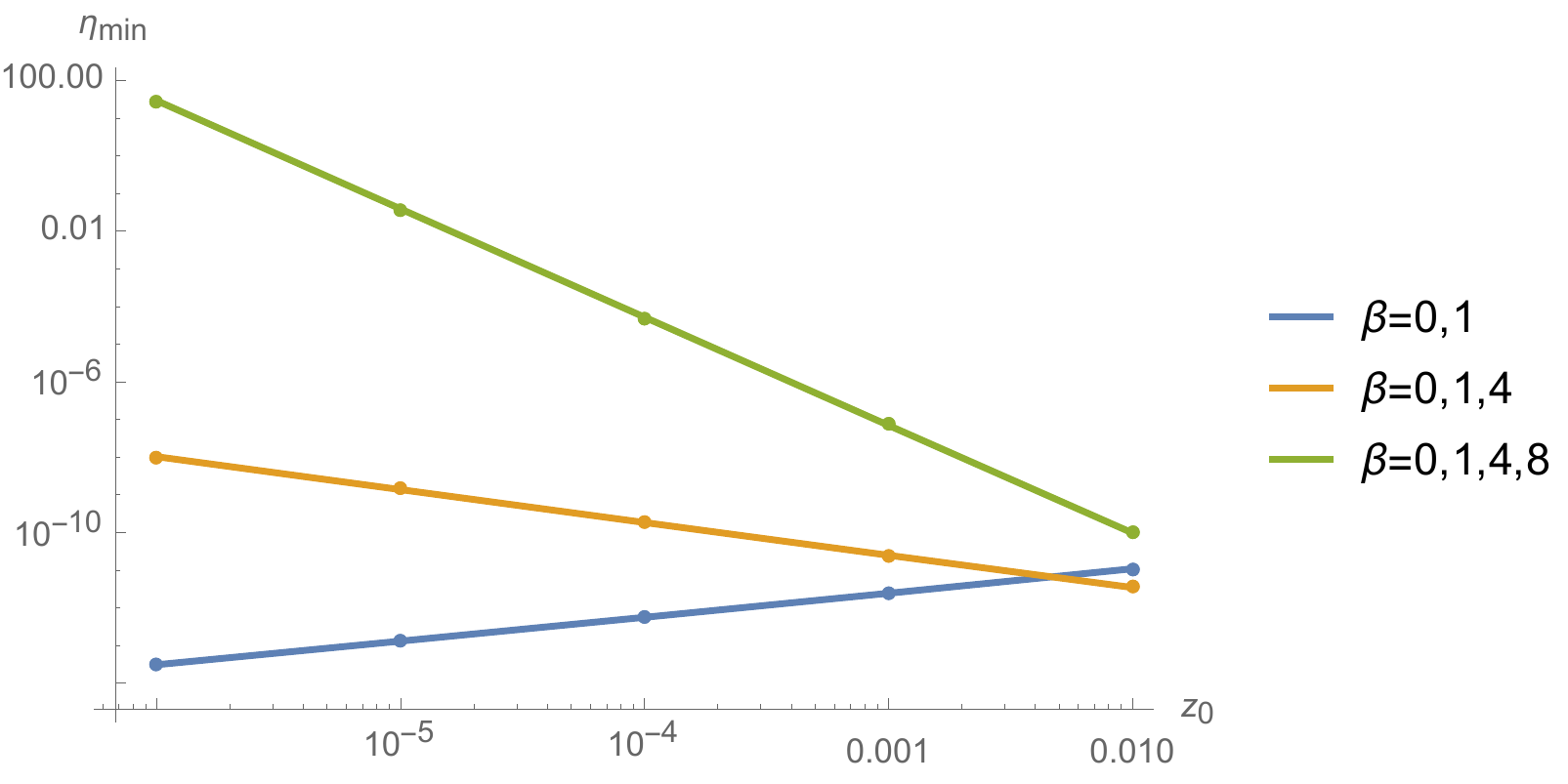}
\caption{Log-log plot of $\eta_\text{min}(z_0)$ with a refined measure \eqref{measure-cutoff}. 
Here the measure depends on the cutoff $\b_\ast=1,4,8$. 
The sampling points are the same as those in Fig. \ref{figure:scaling}. 
The scaling exponents decrease with the $\b$ cutoff and become negative, 
so the $\b$ truncation is not the main source 
and a finite $\eta_\text{min}$ is due to absence of crossing solution. 
}
\label{figure:scaling-refined}
\end{center}
\end{figure}

We can also consider a refined norm with a cutoff dependent measure. 
Near the lightcone, the lowest $\b$ contribution dominates the OPE truncation error, 
so we use 
\be
\mu_\text{refined}(z,\bar z)=\big|z^{\b_\ast/2}-(1-\bar z)^{\b_\ast/2}\big|^{-2}\,,
\label{measure-cutoff}
\ee
where $\b_\ast$ is the cutoff for the $\b$ summation in \eqref{k-sum}. 
If a crossing solution exists, the exponents should always be positive 
because the OPE truncation errors are of higher order in $z, 1-\bar z$ than $\m_\ast^{-1/2}$. 
In Fig. \ref{figure:scaling-refined}, we compare the results of different $\b_\ast$. 
One can see that the exponent $\a$ decreases with the cutoff $\b_\ast$ and becomes negative, 
implying that the OPE truncation is not the main origin of $\eta_\text{min}>0$. 
The approximate values of the scaling exponents are  0.63(1), -0.87(1), -2.87(1), 
where the latter two are consistent with $\a_\text{refined}\approx\a_\text{simple}-\b_\ast/2$. 
A negative exponent also implies a divergent $\eta_\text{min}$ in the double lightcone limit $z,1-\bar z\rightarrow 0$, 
providing a clear signature for absence of crossing solution. 

The $\eta$ minimization results have a geometric interpretation,  
as it induces a special vector $|N\rangle$ orthogonal to the basis vectors. 
The squared minimal distance $\eta^2_\text{min}$ is precisely the inner product of $|N\rangle$ and $|d\rangle$. 
When $\eta_\text{min}>0$, there is no crossing solution  
due to a finite distance between the target point and the space spanned by the basis vectors.

\section{Analytical conformal bootstrap}
For a deeper understanding, let us study the crossing equation \eqref{crossing-compact} in the analytic lightcone expansions. 
We will find obstructions from both regular and bi-singular terms. 

First, we discuss the inconsistency from regular terms. 
Near the double lightcone limit, the target vector can be well approximated by
\be
|d\rangle&=&\frac z {4\sqrt 2}-\frac {3z(1-\bar z)^{\frac 1 8}
} {16\,G}-(z\leftrightarrow 1-\bar z)+\dots\,,
\label{d-explicit}
\ee
where $G=\G(1/4)^2\,(2\pi)^{-3/2}$ is Gauss's constant and $\dots$ indicates higher order terms. 
The scaling behaviour $\eta_\text{min}(z_0)\propto z_0^{1.125}$ in the numerical analysis  
is associated with the leading regular term in \eqref{d-explicit}:
\be
z\,(1-\bar z)^{\frac 1 8}-z^{\frac 1 8}\,(1-\bar z)\,.
\label{d-regular}
\ee
In the lightcone limit $z\rightarrow 0$, the exponents of $z$ are associated with the half twists of primary and descendant states 
in the direct-channel OPE. 
In the double lightcone expansion, we expect that the exponents of $u,v$ are associated with intermediate twists \cite{Alday:2015ota}: 
$v^{\D_\s}\mathcal G(z,\bar z)=\sum_{i,j} c_{i,j}\,u^{\t_i/2}\,v^{\t_j/2}$, such as \eqref{G0}. 
Since the double-twist trajectories $[\s\s]_n$ are absent in 2d, 
the first term $z\,(1-\bar z)^{1/8}$ can only come from the direct-channel contribution. 
One can show that the structure of $k_{\b}(z)$ in \eqref{k-sum} is inconsistent with the explicit expression of $|d\rangle$,  
so the crossing equation \eqref{crossing-compact} has no solution. 

The $d=2$ solution is very special. 
All the regular terms have vanishing coefficients, 
so the double-twist trajectories $[\s\s]_n$ can be absent. 
At first order in $\e=d-2$, the $d$-dependence of conformal symmetry requires the presence of double twist states in the $\s\times\s$ OPE. 
From this analytic perspective, the spectrum transition takes place at $d=2+0^+$. 
Usually, the presence of double-twist states is based on the assumption of a twist gap 
\cite{Fitzpatrick:2012yx, Komargodski:2012ek}, 
but here we show that they are required by conformal symmetry even if the twist gap vanishes. 
Furthermore, we expect the existence of other double/multi-twist states, 
but the more complicated ones should be suppressed by higher powers of $\e$. 

Second, we consider the inconsistency associated with bi-singular terms. 
The presence of double-twist trajectories is not sufficient. 
Another obstruction to solving \eqref{crossing-compact} 
is the large spacing of the twist spectrum \eqref{tau0}. 
We can simplify the analysis by focusing on the bi-singular terms. 
To match the power laws in $|d\rangle_\text{b.s.}$, 
the functions $A_\b(\bar z),B_\b(\bar z)$ in \eqref{k-sum} should take the form
$\sum_{k=0}^\infty \big(a_{0,k}+a_{1,k}\log(1-\bar z)\big)(1-\bar z)^{k/2-1/8}\,,
$ with $a_{n,k}$ replaced by $b_{n,k}$ for $B_\b(\bar z)$. 
We introduce $\log (1-\bar z)$ because $\pa_\b k_{\b}(z)$ involves $\log z$. 
The exponents take the expected values and there is no double-twist exponents, 
so naively we may try to solve the crossing equation order by order. 
Let us count the total power of $z, 1-\bar z$. 
For example, the first line of eq.(10) in Supplemental Material contains terms of order $1, 3/2$.  
After solving the crossing equation to order $2$, we substitute the solutions of $a_{n,k},b_{n,k}$ into the $5/2$ order equation.  
We find that the sum below has a fixed coefficient
\be
&&\Big(\D_\s^{(1)}|\D_\s\rangle+\sum_{i}\D_i^{(1)}|\D_i\rangle+\sum_{i} \l_i^{(1)}|\l_i\rangle
+|d\rangle\Big)_\text{b.s.}
\nn
&=&-\frac {6\sqrt 2}{539}\Big(z\,(1-\bar z)^{3/2}-z^{3/2}\,(1-\bar z)\Big)+\dots\,,
\label{bi-singular-obstruction}
\ee
so the bi-singular part of the crossing equation \eqref{crossing-compact} has no solution beyond order 2. 
To construct a crossing solution, 
one can reduce the spacing of the twist spectrum from 4 to 2, 
as in the standard case of generalized free theory. 

The $d=2$ solution is possible because of the special structure of 2d conformal blocks. 
As 2d global conformal blocks are invariant under $\ell\rightarrow -\ell$, 
the spectrum is symmetric in twist $\D-\ell$ and conformal spin $\D+\ell$. 
This explains the large spacing in intermediate twist spectrum, which is dual to that of $2\ell$ 
\cite{fn18}.
As only even spin states appear in the $\s\times \s$ OPE, the 2d twist spacing is 4. 
This large spacing is inconsistent with the general $d$ structure of conformal blocks. 

\section{Discussion}
We have investigated the $d=2+\epsilon$ critical Ising model using 
novel numerical and analytical conformal bootstrap methods. 
Our analyses of the crossing equation \eqref{crossing-compact} disprove 
the naive expectation that the leading corrections are linear in $\e$. 
The $d$-dependence of global conformal symmetry 
implies the existence of new intermediate states, such as double twist trajectories.   
But the intermediate spectrum is the same as the $d=2$ case at order $\e$ if the naive expectation is correct. 
No solution to the crossing equation can be found 
due to the linear independence of conformal blocks. 
Since the obstructions are related to the speciality of $d=2$, 
we expect them to appear also in other 2d CFTs. 

A direct consequence is that the leading corrections to the 2d data should be more singular than $\e^1$. 
The presence of $\e^a$ corrections with $0<a<1$ will imply strong non-unitarity of the Ising CFT 
below $d=2$. 
This is consistent with the observation of two kinks in the $d<2$ unitary bootstrap bounds in \cite{Golden:2014oqa}. 
The leading corrections may take the form of order $\e^{1/k}$ with $k=2,3,\cdots$. 
One can rule out the possibility that only scaling dimensions receive $\e^{1/k}$ corrections by 
adding higher $\b$ derivatives of $k_{\b}(z)$ to \eqref{k-sum}, 
then there should be infinitely many new states. 
Similar to the XY model results in \cite{Nelson-Fisher}, 
the simplest resolution could be that the leading corrections are of order $\e^{1/2}$. 
For $d<2$, the scaling dimensions can be complex conjugate pairs 
and the OPE coefficients of new states can be imaginary numbers. 
For perturbative RG fixed points, 
the $\e^{1/2}$ behavior has also been found in the cases with two marginal operators, 
such as the $d=4-\epsilon$ random Ising model \cite{Khmelnitskii}. 
In general, the square root behaviour can appear 
around a bifurcation point at which two fixed points collide 
\cite{Gukov:2016tnp,Osborn:2017ucf,Gorbenko:2018ncu,Rychkov:2018vya}.   

Although we show there are infinitely many new states, 
it is still unclear if the low-lying scaling dimensions receive singular corrections.  
It may be helpful to learn from other analytical insights \cite{Rychkov:2015naa,Skvortsov:2015pea,Gliozzi:2016ysv,Gliozzi:2017hni,Gopakumar:2016wkt,Gopakumar:2016cpb,Giombi:2016hkj,Alday:2016jfr,Alday:2017zzv,Carmi:2020ekr,Alday:2016njk, Caron-Huot:2017vep,Carmi:2019cub,Mazac:2019shk,Penedones:2019tng,Caron-Huot:2020adz}. 
It would also be fascinating to study other strongly coupled CFTs in $2+\e$ dimensions. 
For more complex problems, 
it could be useful to assume a hierarchical structure in operator product expansion \cite{Li:2017agi,Li:2017ukc}. 

Many statistical physics models violate reflection positivity.  
Similarly, the boundary/defect bootstrap \cite{Liendo:2012hy}\cite{Gliozzi:2015qsa}
\cite{Billo:2016cpy,Gadde:2016fbj,Liendo:2016ymz,Lemos:2017vnx} 
and 
thermal bootstrap \cite{El-Showk:2011yvt,Iliesiu:2018fao,Iliesiu:2018zlz} problems do not obey positivity constraints. 
In the usual numerical bootstrap, 
the positivity constraints are crucial to the derivation of rigorous bounds. 
Here we show that the inconsistent theory space can be ruled out without using positivity. 
We plan to revisit the non-positive bootstrap problems from the new perspective.  

\begin{acknowledgments}
I would like to thank Hugh Osborn, Slava Rychkov, Ning Su and 
other oganizers/participants of Bootstat 2021 for enlightening comments. 
I am particularly grateful to Slava Rychkov for valuable comments on the draft. 
The Bootstat 2021 program took place at 
Institut Pascal at Université Paris-Saclay with the support of the program “Investissements d’avenir” ANR-11-IDEX-0003-01. 
This work was partly supported by the 100 Talents Program of Sun Yat-Sen University, 
Okinawa Institute of Science and Technology Graduate University (OIST) 
and JSPS Grant-in-Aid for Early-Career Scientists (KAKENHI No. 19K14621).
\end{acknowledgments}

\bibliographystyle{JHEP}

\begin{thebibliography}{10}
\bibitem{fn1}
There are logarithmic corrections at $d=4$ \cite{4d-log}.
\bibitem{fn4}
It is well known that the 2d Ising model can be mapped to a free fermion theory. 
It would be interesting to extend the fermionization to $2+\e$ dimensions.
\bibitem{fn9}
A simple example is: 
$a_0+a_2\, t^2=b_0+b_1\,t^1+b_2\,t^2$, 
which has no solution that is valid at generic $t$ if $b_1\neq 0$.
\bibitem{fn14}
In fact, $A_\b(\bar z), B_\b(\bar z)$ should satisfy additional constraints,  
but the enlarged space has no solution as well.
\end{thebibliography}

\end{document}